\documentclass[letter]{article}

\usepackage[margin=1in]{geometry}
\usepackage{tikz}
\usetikzlibrary{arrows}
\usepackage{amssymb}
\usepackage{amsmath}
\usepackage[super,comma,sort&compress,numbers]{natbib}
\usepackage{bm}
\usepackage{array,multirow}
\usepackage{hyperref}
\usepackage{enumerate}
\usepackage[inline]{enumitem}
\usepackage{setspace}
\usepackage[document]{ragged2e} 
\usepackage{sectsty}
\usepackage{titlecaps}
\usepackage{authblk} 

\newcommand{\bei}{\begin{itemize}}
\newcommand{\eei}{\end{itemize}}
\newcommand{\beq}{\begin{equation}}
\newcommand{\eeq}{\end{equation}}
\newcommand{\ben}{\begin{enumerate}}
\newcommand{\een}{\end{enumerate}}

\newcommand{\argmin}{{\rm arg\,min}}

\DeclareMathOperator*{\E}{E}
\DeclareMathOperator*{\V}{V}

\DeclareMathOperator*{\expit}{expit}

\title{Sensitivity Analysis for Unmeasured Confounding via Effect Extrapolation}

\author[1]{Wen Wei Loh}
\affil[1]{Department of Data Analysis, Ghent University, Gent, Belgium.}
\author[2,3]{Stijn Vansteelandt}
\affil[2]{Department of Applied Mathematics, Computer Science and Statistics, Ghent University, Ghent, Belgium}
\affil[3]{Department of Medical Statistics, London School of Hygiene and Tropical Medicine, London, United Kingdom}

\begin{document}

\maketitle
\sectionfont{\large\MakeUppercase} 

\begin{abstract}
Inferring the causal effect of a non-randomly assigned exposure on an outcome requires adjusting for common causes of the exposure and outcome to avoid biased conclusions. Notwithstanding the efforts investigators routinely make to measure and adjust for such common causes (or confounders), some confounders typically remain unmeasured, raising the prospect of biased inference in observational studies. Therefore, it is crucial that investigators can practically assess their substantive conclusions' relative (in)sensitivity to potential unmeasured confounding. In this article, we propose a sensitivity analysis strategy that is informed by the stability of the exposure effect over different, well-chosen subsets of the measured confounders. The proposal entails first approximating the process for recording confounders to learn about how the effect is potentially affected by varying amounts of unmeasured confounding, then extrapolating to the effect had hypothetical unmeasured confounders been additionally adjusted for. A large set of measured confounders can thus be exploited to provide insight into the likely presence of unmeasured confounding bias, albeit under an assumption about how data on the confounders are recorded. The proposal's ability to reveal the true effect and ensure valid inference after extrapolation is empirically compared with existing methods using simulation studies. We demonstrate the procedure using two different publicly available datasets commonly used for causal inference.

\end{abstract}


\section*{Introduction}

When inferring the causal effect of a non-randomly assigned (point) exposure on an outcome, common causes (i.e., {\em confounders}) of both exposure and outcome that are unadjusted for can induce associations that lead to biased estimates.
Investigators therefore strive to measure as many (pertinent) confounders as possible when analyzing data from observational studies in efforts to eliminate such biases due to unmeasured confounding. 
But in most realistic settings, some confounders remain unadjusted for, raising the prospect of such biases being merely reduced, but not entirely eliminated. 
For example, there may be either confounders that are unfeasible or impractical to observe or record \citep{Rosenbaum:2002}, or (measured) confounders excluded by routine variable selection techniques \citep{wuthrich2019omitted}.
It is therefore crucial that investigators can practically assess the relative stability of their conclusions to dormant confounding that remains unadjusted for.

\section*{Proposed sensitivity analysis strategy}

In this paper, we propose a novel sensitivity analysis strategy to assess how different the conclusions would be after removing unmeasured confounding. To understand the proposal, consider the following thought experiment based on an illustration presented later. Suppose that you were given data from a simple randomized trial on a (dichotomous) AIDS therapeutic treatment, an outcome measuring patients' CD4 T cell count, and a collection of baseline covariates. However, suppose that you were not told that the data originated from a randomized experiment. Then upon analyzing the data, you would observe that regardless of the covariates adjusted for, roughly the same average treatment effect estimate is obtained. The relative stability of the effect estimates, across different adjustment sets, would allow you to gain confidence that the data might have originated from a randomized experiment. This confidence would grow as more and more covariates (beyond the initial collection) are subsequently recorded, and you keep on observing a stability in the average treatment effect estimate that is maintained over larger and larger adjustment sets. 

What we learn from the above thought experiment is that a large set of measured covariates can be exploited to provide insight into the likely presence of unmeasured confounding bias, albeit under an assumption about how the confounders are recorded. In particular, we assume that under repeated sampling, covariates are chosen (from a possibly infinitely large collection) according to some sampling mechanism, such that each confounder that simultaneously influences exposure and outcome has a positive (non-zero) probability, or likelihood, of being recorded (and adjusted for). 
Confounders that are unadjusted for would thus induce unmeasured confounding, and result in biased effects. But if that sampling mechanism is well understood, then the bias due to unmeasured confounding can be eliminated, by extrapolating the observable behavior of the effects adjusting for the measured confounders to recover the (true) effect had the entire collection of confounders been adjusted for.
Indeed, suppose for instance that the measured confounders form a random subset of all confounders, where each confounder has an equal probability of being recorded. Our proposal mimics the sampling mechanism as follows. Starting with the set of measured confounders, randomly select a confounder and eliminate it from adjustment so that it can be considered to be unmeasured. Assuming a sufficiently large sample size so that sampling variability can be ignored, calculate the (population-averaged) exposure effect adjusting for only the retained confounders.
Repeat this process of intentionally eliminating from adjustment a single confounder one at a time, until no confounders are adjusted for. 
We can now probe how the (biased) effects change with different amounts of unmeasured confounding; furthermore, we can extrapolate to the (unbiased) effect adjusting for the entire collection of confounders. In practice, however, because the sampling mechanism through which confounders are recorded for adjustment is less well understood, we propose a more heuristic approach to approximate the (non-random) process for recording confounders. 

\section*{Sensitivity analysis via effect extrapolation}

\subsection*{Intentionally eliminating measured covariates from adjustment}

The proposal proceeds by eliminating measured covariates one at a time to intentionally induce unmeasured confounding. Ideally, the order of elimination should be determined using information about the investigators' process for recording confounders. But such information may be difficult to precisely quantify in practice. In this paper, we will thus proceed under the assumption that the covariates exerting the strongest impact on the effect are prioritized to be recorded (and adjusted for) first. We propose a data-adaptive approach to mimic such a process. For simplicity we will initially ignore any sampling variability. 
Starting with the set of measured covariates, eliminate the covariate that causes the smallest change in the (population-averaged) exposure effect when unadjusted for, so that the covariate can be considered as unmeasured. Repeat this until no covariates remain for adjustment. 
Under this assumed process, covariates that are eliminated earlier have lower priority for confounding adjustment and thus the weakest impact on the effect. 

To aid visualizing the impact of unmeasured confounding on the effect, we partition the space of all possible covariate subsets into orbits \citep{crainiceanu2008adjustment}, where the $j$-th orbit comprises all subsets with $j+1$ covariates, including an intercept.
Let $J$ denote the total number of measured covariates so that there are $J+1$ different orbits. 
We briefly introduce the notation as follows.
In a sample of size $n$, for individual $i=1,\ldots,n$, denote the binary exposure by $A_i$ and the outcome of interest by $Y_i$.
Let $Y^a_i$ denote the potential outcome for $Y_i$ if, possibly counter to fact, individual $i$ had been assigned to exposure level $A_i=a$. In this paper, our interest is in the marginal exposure effect, defined as $\psi=\E(Y^1)-\E(Y^0)$. The first part proceeds by repeating the following steps for each orbit indexed by $j=J,\ldots,1$.
\begin{enumerate}
\item Let $L^{j+1}$ denote the subset of covariates remaining in the $(j+1)$-th orbit. When $j=J$, let $L^{J+1}$ denote the full set containing all measured covariates and the constant (intercept) $1$. Denote each of the covariates in $L^{j+1}$ by $L^{j+1, k}, k=1,\ldots,j$.
These $j$ covariates are therefore candidates to be eliminated from confounding adjustment in the $j$-th orbit. 
\item 
For $k=1,\ldots,j$, calculate the exposure effect estimator conditional on the covariates in $L^{j+1}$ excluding the single covariate $L^{j+1, k}$, i.e., $(L^{j+1} \setminus L^{j+1, k})$, which we denote simply by $\widehat \psi_{j+1,k}$.
\item Let $\psi_{j+1,k} = \E(\widehat \psi_{j+1,k})$ denote the expected value of the effect estimator, and $\psi_{j+1} = \E(\widehat \psi_{j+1})$ denote the expected value adjusting for the covariates in $L^{j+1}$. We will first explain how to proceed when the true values of the (asymptotic) expectations $\psi_{j+1,k}$ and $\psi_{j+1}$ are known, so that the true bias caused by each candidate confounder being unadjusted for can be determined exactly. Let $k^\ast$ denote the index of the candidate confounder that yields the smallest (squared) magnitude of the following difference: 
\beq\label{eq:forwardselect_minimin}
k^\ast = \underset{k}{\argmin} \left(\psi_{j+1,k}-\psi_{j+1}\right)^2.
\eeq
Define the subset of remaining covariates in the $j$-th orbit to be $L^j=(L^{j+1} \setminus L^{j+1, k^\ast})$.
Denote the effect estimator that adjusts for only the (retained) confounders in $L^j$ by $\widehat \psi_{j} = \widehat \psi_{j+1,k^\ast}$.
\end{enumerate}
Repeating the steps above for $j=J,\ldots,1$, returns a sequence of (nested) covariate subsets $L^{J+1} \supset \ldots \supset L^1$, where a single, different measured covariate is dropped in each orbit, relative to those in the previous (larger) orbit. Each subset indexes an effect so that the sequence of effect estimators, $\widehat\psi_{J+1}, \ldots, \widehat\psi_{1}$, quantifies the impact of unmeasured confounding induced by eliminating covariates from adjustment in turn. 

The assumption that $\psi_{j+1,k}$ and $\psi_{j+1}$ are known when evaluating the criterion \eqref{eq:forwardselect_minimin} can only be considered to hold approximately in very large samples where sampling error can be ignored. Otherwise, estimators are needed to approximate \eqref{eq:forwardselect_minimin}.
Asymptotically unbiased estimators may nonetheless suffer from inaccuracies in finite samples due to sampling variability, so that merely plugging in the (sample) estimators for the (population) effects can produce incorrect results. To see why, consider a confounder that is strongly associated with exposure, but weakly associated with outcome, so that it causes only a small (true) bias when unadjusted for. 
But adjusting for this confounder reduces the precision of the effect estimator \citep{brookhart2006variable}, so that the observable change in estimated effect may be deceptively large relative to the true bias when the confounder is eliminated, simply due to sampling variability. 
The sampling uncertainty of the effect estimators can be acknowledged to more accurately assess the true biases due to unmeasured confounding as follows. Let $\widehat\psi_{j}$ and $\widehat\psi_{j^\prime}$ denote the effect estimators from two different orbits, e.g., $j$ and $j^\prime$. The expectation of the squared difference between the estimators can thus be decomposed as:
\begin{align*}
\E\left\{\left(\widehat\psi_{j}-\widehat\psi_{j^\prime}\right)^2\right\}
&=\E\left[\left\{\widehat\psi_{j}-\widehat\psi_{j^\prime}-(\psi_{j}-\psi_{j^\prime})+(\psi_{j}-\psi_{j^\prime})\right\}^2\right] \\
&=\V\left(\widehat\psi_{j}-\widehat\psi_{j^\prime}\right) + (\psi_{j}-\psi_{j^\prime})^2 
+ 2\E\left[ \left\{\widehat\psi_{j}-\widehat\psi_{j^\prime}-(\psi_{j}-\psi_{j^\prime})\right\}(\psi_{j}-\psi_{j^\prime})\right] \\
&=\V\left(\widehat\psi_{j}-\widehat\psi_{j^\prime}\right) + (\psi_{j}-\psi_{j^\prime})^2,
\end{align*}
where $\V(X)$ denotes the asymptotic variance of $X$.
The last equality follows from the asymptotic unbiasedness of $\widehat\psi_{j}$ and $\widehat\psi_{j^\prime}$.
By considering a squared difference between the effects in \eqref{eq:forwardselect_minimin}, the squared difference between the corresponding estimators is no longer an unbiased estimator, with bias proportional to the (population) variance of the difference between the effect estimators.
The ``debiased'' squared difference between the effects is thus $(\psi_{j}-\psi_{j^\prime})^2 = \E\left\{\left(\widehat\psi_{j}-\widehat\psi_{j^\prime}\right)^2\right\} - \V\left(\widehat\psi_{j}-\widehat\psi_{j^\prime}\right)$. 
In practice, values of \eqref{eq:forwardselect_minimin} can be calculated by plugging in the sample estimator for the population mean, and the (asymptotic) variance estimator for the population variance; i.e.,
$\left(\widehat\psi_{j+1, k}-\widehat\psi_{j+1}\right)^2-\widehat\V\left(\widehat\psi_{j+1,k}-\widehat\psi_{j+1}\right)$.
When this quantity is non-positive in practice, we will simply set its value to zero; when there are multiple covariates with the same zero value, we will randomly select a covariate among these to be eliminated.
The sampling variability of the estimates is thus recognized (and removed) when assessing the potential bias caused by eliminating each candidate confounder from adjustment. 
Because these sample approximations can be imperfect, we defer to future work investigating other (computationally-intensive) methods, such as cross fitting, to more precisely estimate the true changes in the population-level effects between orbits.

To facilitate comparing effect estimators that condition on different confounders across orbits when evaluating non-collapsible effect measures \citep{greenland1999confounding}, we recommend focusing on the same (marginal) estimand to avoid falsely interpreting changes in the estimated effects due to non-collapsibility as a result of the eliminated confounders' influence on the effect.
We will therefore employ an effect estimator based on doubly robust standardization \citep{vansteelandt2011invited} to ensure that the selected sequence of confounders is determined using a collapsible measure.
The differences between the exposure effect estimators, and the associated variances, can be consistently estimated under settings with (non-)linear parametric regression models for the exposure and the outcome. 
This approach delivers an unbiased estimator if either the outcome or the exposure model is correctly specified, without amplifying biases that may arise due to the misspecified model. 
Furthermore, the (asymptotic) variance estimator needed to calculate \eqref{eq:forwardselect_minimin} can be derived in closed form for computational efficiency. We describe how to calculate this quantity, including the effect estimator in a given orbit, in Appendix~\ref{sect:DRestimator}.
In principle, any criterion can be used in place of \eqref{eq:forwardselect_minimin} to eliminate the confounders in turn toward mimicking (or realizing) the process of recording confounders for adjustment. For example, theoretical background or study design knowledge, or empirical covariate prioritization or importance measures \citep{Loh2020stability,williamson2020}, can be exploited.

\subsection*{Accounting for sampling uncertainty in the effect estimators}\label{sect:perturbing}

We have sought to construct the (ideal) sequence in which covariates are eliminated from adjustment in turn using the true causal effects in \eqref{eq:forwardselect_minimin}. But the inherent sampling uncertainty of the estimators only permits ``guessing'' what the sequence should be. To acknowledge the uncertainty, we will give the data several opportunities to approximate the (true) process of prioritizing covariates for adjustment, by {\em perturbing} the sequence to express our uncertainty about the effects while investigating unknown confounding.

We describe how to construct a perturbed sequence by modifying steps 2 and 3 of the proposed procedure, and defer technical details to Appendix~\ref{sect:DRestimator}. In step 2, for each candidate confounder indexed by $k$, determine the maximum likelihood estimates (MLEs) of the (coefficients in the) exposure and outcome models used to calculate the effect estimator $\widehat \psi_{j+1,k}$. Then randomly draw a value of the estimated outcome model coefficients from their joint sampling distribution. Calculate the perturbed effect estimator using the randomly drawn coefficient values, which we denote by $\tilde \psi_{j+1,k}$. 
Hence $\tilde \psi_{j+1,k}$ varies from $\widehat \psi_{j+1,k}$ on the basis of sampling uncertainty, where the former is based on a random draw from the estimated coefficients' sampling distribution, and the latter is based on the MLE of the coefficients. In step 3, the confounder to be eliminated from adjustment can then be determined using the perturbed estimators $\tilde \psi_{j+1,k}$, and $\tilde \psi_{j+1}$, in place of $\widehat \psi_{j+1,k}$, and $\widehat \psi_{j+1}$; in other words, set 
$k^\ast = \underset{k}{\argmin} \left(\tilde\psi_{j+1, k}-\tilde\psi_{j+1}\right)^2-\widehat\V\left(\tilde\psi_{j+1,k}-\tilde\psi_{j+1}\right)$.
The values of the resulting sequence of perturbed effect estimators, $\tilde \psi_{J+1}, \ldots, \tilde \psi_{1}$, will thus differ from the observed sequence (which is based on the MLEs), even when the covariates are eliminated in the same order.
A Monte Carlo sampling distribution can be readily obtained by constructing e.g., $B = 500$ perturbed sequences.

Ideally, when the precise probability model for recording confounders is known, the perturbed effect estimators should be constructed using random draws from the implied joint sampling distribution of the model coefficients across all orbits. The reduced variability of the effect estimators due to the sequence for eliminating covariates being known would likely yield a trajectory that fluctuates less between orbits. A Bayesian approach may facilitate estimating such a sampling distribution, e.g., in terms of the (marginal) posterior probabilities of adjusting for each covariate. Because we have focused on the conceptual development of the proposed sensitivity analysis in this paper, more complex approaches to sample sequences of effect estimators from their joint distribution are deferred to future work.

\subsection*{Extrapolating to the effect adjusting for unmeasured confounders}\label{sect:extrapolate}

In the second part of the proposed strategy, we extrapolate each unique sequence of (biased) effect estimators to the predicted (unbiased) effect, had additional hypothetical confounders been further adjusted for. The assumption that there exists a (possibly infinitely) large collection of covariates, among which only a subset is revealed for confounding adjustment, and no covariate is precluded from being recorded (and adjusted for) under some unknown sampling mechanism is therefore necessary.
In particular, under repeated observed samples we may each time observe a different set of measured covariates (where some are possibly sampled with unit probability, but none are sampled with zero probability), so that in the long run, across many repeated samples, even the weakest confounders (that are most likely to remain unmeasured) can be adjusted for. 

We will fit a natural cubic spline to each (perturbed) sequence of exposure effect estimators, e.g., $\widehat \psi_{j}, j=1,\ldots,J+1$, with the number of measured covariates adjusted for ($0, \ldots, J$, after excluding the intercept) as a predictor. Natural cubic splines permit flexibly evaluating the trajectory of the exposure effect estimator as unmeasured confounding is reduced one covariate at a time. The predicted value of the effect estimator, had a given number of hypothetical confounders been further adjusted for, can then be (linearly) extrapolated to. Suggestions on fitting natural cubic splines in practice are described in Appendix~\ref{sect:extrapolate_naturalcubicsplines}.
This second part of the strategy is inspired by the Simulation Extrapolation (SIMEX) approach \citep{cook1994simulation} for different settings with measurement error in regression predictors. 

To express uncertainty, a $100(1-\alpha)\%$ Wald confidence interval (CI) may be constructed for each sequence of effect estimators; i.e., $\widehat \psi_{j} \pm z_{1-\alpha/2} \sqrt{\widehat\V(\widehat \psi_{j})}, j=1,\ldots,J+1$, where $z_{1-\alpha/2}$ is the $1-\alpha/2$ percentile of a standard normal distribution. Separate natural cubic splines can then be fitted to each endpoint of the CI, and the endpoints similarly extrapolated to a given number of additional hypothetical confounders.
Repeating the extrapolations for each perturbed sequence yields $B$ different extrapolated $100(1-\alpha)\%$ CIs.
For a given number of hypothetical confounders, the extrapolated CIs across all perturbed sequences (including the observed sequence) can be combined into an {\em uncertainty interval}.
Let the lower (upper) endpoint of the uncertainty interval be the $2.5$ (or $97.5$) percentile among the lower (or upper) endpoints of all the extrapolated CIs. Using the (symmetric) 95 percentiles reduces the uncertainty intervals' susceptibility to (i) perturbations that yield extreme CIs, and (ii) overly-conservative conclusions, as compared to using the minimum (and maximum) of the extrapolated CIs.
We show empirically that the uncertainty intervals cover the true effect either at, or above, the nominal level of the constituent CIs under a variety of settings in simulation studies; details are described in Appendix~\ref{sect:sims}. 

Instead of extrapolating to a given (additional) number of hypothetical confounders, another possibility is to extrapolate to the (smallest) number of confounders for the uncertainty interval to either include or exclude zero so that the statistical significance of the effect estimate changes. The extent of unmeasured confounding required to change the conclusions can thus be quantified in terms of the number of hypothetical (unmeasured) confounders, rather than the strength of a single hypothetical confounder. We develop this interpretation using the illustrations in the next section.

\section*{Illustrations}\label{sect:examples}

We first elaborate on the thought experiment in the introduction to demonstrate how the proposal can reveal a stability in the effect when unmeasured confounding is absent. We use data from an AIDS randomized clinical trial, and defer details to Appendix~\ref{sect:ACTG175}. Suppose that knowledge about treatment being randomized, so that confounding of the exposure-outcome relation was unlikely, was hidden from us. Adjusting for different covariates did not greatly affect the exposure effect estimates, as shown by the relatively stable trajectory in Figure~\ref{fig:plot-speff2}. Across all perturbations, the 95\% CIs for the exposure effect excluded zero, even when none of the measured covariates were adjusted for. Based on the extrapolations, adjusting for additional hypothetical confounders appeared unlikely to yield an uncertainty interval that would include zero, suggesting an insensitivity to hypothetical unmeasured confounding.

We next apply the proposal to an observational study on the effectiveness of Right Heart Catheterization (RHC) in the initial care of critically ill patients \citep{connors1996effectiveness}.
The exposure variable was defined to be whether or not a patient received an RHC within 24 hours of admission. A binary outcome was defined based on whether a patient died at any time up to 180 days since admission. We considered a reduced dataset with 2707 participants having complete data on 72 covariates (one covariate that was singular in the reduced dataset was dropped), so that the exposure and outcome models with all covariates can be fitted. 

\begin{figure}[!ht]
    \centering
\includegraphics[width=\linewidth,keepaspectratio]{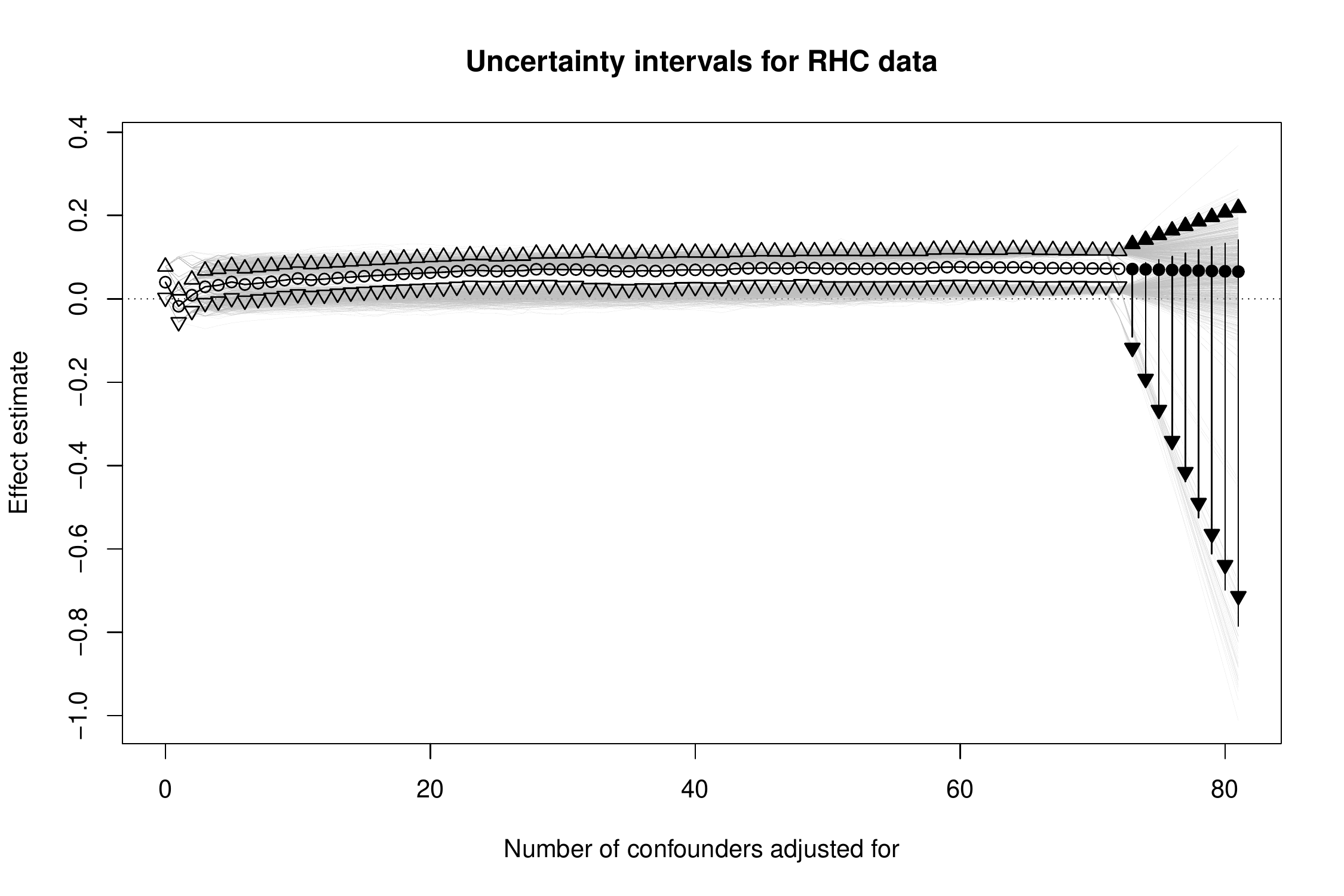}
    \caption{
    Estimators of the average exposure effect from each orbit, for the RHC data. 
Circles indicate the point estimates, and (inverted) triangles indicate the (lower) upper endpoint of the 95\% CI.
Empty points indicate the sequence of effect estimates constructed by eliminating a single measured confounder from adjustment in turn. The curve corresponds to a fitted natural cubic spline.
Filled points indicate the (extrapolated) predictions after adjusting for $q$ additional unmeasured confounders. 
Each thin grey line corresponds to the natural cubic spline fitted to a perturbed sequence. 
The vertical lines indicate the symmetric 95 percentiles of the predicted effects across all $B$ perturbations.
    \label{fig:plot-rhc}}
    \end{figure}

The sequence of effect estimators constructed by eliminating covariates one at a time following the criterion in \eqref{eq:forwardselect_minimin} is plotted (as empty circles) in Figure~\ref{fig:plot-rhc}. 
Adjusting for all the measured covariates yielded an estimated effect of $0.07$, with a 95\% CI of $(0.03, 0.11)$; the estimates were relatively stable above zero even when (a few) covariates were eliminated.
We calculated $B=500$ perturbed sequences of the estimators and their corresponding 95\% CIs. 
Natural cubic splines were fitted to each (perturbed) sequence of the estimators and endpoints of the CI, and the predicted effects, and uncertainty intervals, extrapolated to.
Almost all the predicted effects across the perturbations remained positive after adjusting for unmeasured confounding; e.g., 4\% (8\%) of the predicted effects were negative after adjusting for one (nine) hypothetical confounder(s). The high sampling uncertainty in the predicted effects can be observed from their 95 percentiles (vertical lines) being wider than the endpoints of the uncertainty intervals (filled inverted triangles) in Figure~\ref{fig:plot-rhc}. 
Furthermore, the uncertainty intervals included zero with only one hypothetical unmeasured confounder. 
The conclusion of a (statistically significant) positive effect adjusting for only the measured covariates is thus likely to change to an effect indistinguishable from zero after accounting for unmeasured confounding.
Our findings are line with \citet{Lin1998}: in spite of the efforts made by the study investigators to ascertain and adjust for all the known risk factors, the conclusion that there was a (barely statistically significant) harmful exposure effect of RHC is sensitive to unmeasured confounding.

\section*{Comparisons with existing approaches}

Existing sensitivity analyses for unmeasured confounding typically encode the association between a single hypothetical confounder with either the exposure or the outcome (or both) as (separate) sensitivity parameters \citep{Rosenbaum1987,imbens2003}.
Because the observed data do not identify the sensitivity parameters, different ``reference'' values are specified, typically by selecting a single measured confounder (among possibly many) as a proxy for the hypothetical confounder, then judging the (minimum) extent that inference is affected \citep{carnegie2016assessing,ding2016sensitivity,dorie2016flexible,veitch2020sense}.
We defer reviewing (other) existing methods to Appendix~\ref{sect:review_existing}. 
In contrast to these approaches, we exploit the richness of all available joint information on the measured covariates, by probing the effects across different amounts of intentionally induced unmeasured confounding using different adjustment sets, without being limited to particular interpretations and values of the (unidentifiable) sensitivity parameters. 
This can turn out to be informative. For example, suppose that a researcher has recorded all confounders so that there truly exists no unmeasured confounding, but this fact is unknown to them. Existing methods would nonetheless imply that unmeasured confounding remains a possibility in the worst-case scenario, which may inflate the degree of uncertainty using subjective judgment \citep{10.1093/ije/dyaa095}. 
In contrast, our proposal exploits features of the available data and study design to better inform the plausibility for unmeasured confounding.
In particular, if all confounders are indeed adjusted for, and additional ``redundant'' covariates that are associated with only either exposure or outcome are subsequently recorded (\citet{10.1093/ije/dyaa095} describes such an example \citep{brown2017association}), then our proposed analysis would reveal little or no sensitivity of inference due to unmeasured confounding.
We acknowledge the heuristic nature of our proposal, but argue that all sensitivity analyses for unmeasured confounding are inherently heuristic, because the observed data does not identify the causal effect under such settings; see e.g., \citet{10.1093/ije/dyaa127} and the ensuing discussions.

Our proposal therefore differs from existing approaches in a few aspects. 
\begin{enumerate*}[label=(\roman*)]
\item No (unidentifiable) sensitivity parameters are required.
\item No assumptions about the (arbitrary) distribution or scale of the hypothetical confounder(s), or whether they amplify or nullify the effect estimate, need to be imposed.
\item The exposure and outcome can be continuous or non-continuous, so that non-linear models can be readily accommodated for estimating the exposure effect.
\item The sensitivity of exposure effects to the possibility of (un)measured confounding can be concisely inspected using both graphical and numerical methods, as we demonstrated in the illustrations. 
\item The proposed strategy is not limited to the average exposure effect, and can be readily applied to assess the sensitivity of any (scalar) causal effect to unmeasured confounding, such as the effect of exposure among the (un)treated. In principle, the strategy can be readily generalized in future work to more complex settings, such as heterogeneous exposure effects and nonparametric estimation methods.
\item In contrast to existing approaches, the proposal works best when a large collection of covariates has been recorded, and background or empirical information on both measured and unmeasured confounders is available, so that their joint influence on the exposure effect can be meticulously examined.

\end{enumerate*}

\bibliographystyle{abbrvnat}
\bibliography{sensitivity}

\clearpage

\appendix

\section{Exposure effect estimator}\label{sect:DRestimator}
The effect estimator based on doubly robust standardization \citep{vansteelandt2011invited} is calculated as follows.
For a binary exposure, denote the (non-linear) exposure model, conditional on the remaining confounders in the $j$-th orbit $L^{j}$, by $\E(A_i|L^{j}_i)=\Pr(A_i=1|L^{j}_i)=\expit(\alpha_{j} L^{j}_i)$, where the subscript $i$ of $L^{j}_i$ denotes individual $i$ and $\expit(x) = \exp(x)/\{1+\exp(x)\}$.
Define the {\it inverse probability of exposure weight} \citep{rosenbaum1987model} for individual $i$ as:
\beq\label{eq:IPTW_def}
W_i^j = \dfrac{A_i}{\Pr\left(A_i=1|L_i=L^j_i\right)} + \dfrac{1-A_i}{1-\Pr\left(A_i=1|L_i=L^j_i\right)}.
\eeq
The weight $W_i^j$ is the reciprocal of the conditional probability of individual $i$ being assigned the observed exposure $A_i$ given the confounders $L^j_i$. 
Let $\widehat W_i^j$ denote the estimated weights obtained by substituting the maximum likelihood estimators (MLE) for $\alpha_{j}$ in the exposure model.
Fit the outcome regression model $\E(Y|A,L^{j}) = h^{-1}(\psi_{j}^\ast A + \beta_{j} L^{j})$ to the observed data using the aforementioned weights, where $h^{-1}$ is the inverse of a link function $h$.
(The $\ast$ superscript indicates a conditional effect that may differ from the marginal effect $\psi_j$.)
Let $\widehat\E(Y|A,L^{j})$ denote the fitted outcome model obtained by plugging in the MLE for $\psi_{j}^\ast$ and $\beta_{j}$.
A doubly robust estimator of the average potential outcome $\E(Y^a)=n^{-1}\sum_{i} Y^a_i$ for $a=0,1$, is then:
\beq
\widehat\E(Y^a) =
n^{-1} \sum_{i} {\rm I}\{A_i=a\}\widehat W_i^j \left\{Y_i - \widehat\E(Y|A=A_i,L^{j}_i)\right\} + 
\widehat\E(Y|A=a,L^{j}_i),
\eeq
where ${\rm I}\{B\}=1$ if $B$ is true and $0$ otherwise.
The estimator for the marginal exposure effect $\psi=\E(Y^1)-\E(Y^0)$ in the $j$-th orbit is therefore:
\beq\label{eq:DRestimator_obs}
\widehat\psi_{j} = 
n^{-1} \sum_{i} (2A_i-1) \widehat W_i^j \left\{Y_i - \widehat\E(Y|A=A_i,L^{j}_i)\right\} + 
\widehat\E(Y|A=1,L^{j}_i) - \widehat\E(Y|A=0,L^{j}_i).
\eeq
When both exposure and outcome models are correctly specified, an asymptotic expansion around $\psi$ yields the so-called ``influence function'' for individual $i$ as:
\beq
\phi^{j}_i = (2A_i-1) W_i^j \left\{Y_i - \E(Y|A=A_i,L^{j}_i)\right\} + 
\E(Y|A=1,L^{j}_i)-\E(Y|A=0,L^{j}_i) - \psi.
\eeq
Let $\widehat\phi^{j}_i=(2A_i-1) \widehat W_i^j \left\{Y_i - \widehat\E(Y|A=A_i,L^{j}_i)\right\} + 
\widehat\E(Y|A=1,L^{j}_i) - \widehat\E(Y|A=0,L^{j}_i) - \widehat\psi_{j}$ denote the estimated influence function, obtained by plugging in the maximum likelihood estimators for the coefficients in the exposure and outcome models, and substituting the population expectation with a sample average.
The variance of the difference between effect estimators from two different orbits, e.g., $j$ and $k$, is consistently estimated by the sample variance (denoted by $\widehat\V$) of the corresponding difference in estimated influence functions:
\beq\label{eq:var_diff_infn}
\widehat\V\left\{n^{1/2}\left(\widehat\psi_{j}-\widehat\psi_{k}\right)\right\}
=
(n-1)^{-1} \sum_i \left(\widehat\phi^{j}_i - \widehat\phi^{k}_i\right)^2.
\eeq
Consistency and asymptotic normality of the standardized difference \eqref{eq:forwardselect_minimin} with mean zero and variance one directly follow from the law of large numbers and the central limit theorem.
The squared difference between the (asymptotic) expectations of the effect estimators is then approximated by
$\left(\widehat\psi_{j}-\widehat\psi_{k}\right)^2 - \widehat\V\left(\widehat\psi_{j}-\widehat\psi_{k}\right)$.


A perturbed effect estimator can be calculated as follows.
In step 2 of the proposed procedure, in place of calculating the effect estimator $\widehat \psi_{j+1,k}$ for each candidate confounder based on the maximum likelihood estimates (MLEs) of the (coefficients in the) exposure and outcome models, carry out the following steps instead:
\ben[label=2(\alph*)]
\item Fit the exposure model $\Pr(A_i=1|L^{j+1, -k}_i)=\expit(\alpha_{j+1,-k} L^{j+1, -k}_i)$ to the observed data, where $L^{j+1, -k}_i = (L^{j+1}_i \setminus L^{j+1, k}_i)$ denotes the confounders in $L^{j+1}$ excluding the single confounder $L^{j+1, k}$ for individual $i$, and $\alpha_{j+1,-k}$ denotes the corresponding coefficients. Calculate the MLEs, denoted by e.g., $\widehat\alpha_{j+1,-k}$, and the observed weights as defined in \eqref{eq:IPTW_def}, by substituting $\widehat\alpha_{j+1,-k}$ for $\alpha_{j+1,-k}$ in the exposure model.

\item Fit the outcome regression model $\E(Y|A,L^{j+1, -k}) = h^{-1}(\psi_{j+1,-k}^\ast A + \beta_{j+1,-k} L^{j+1, -k})$ to the observed data using the observed weights from the previous step, where $h^{-1}$ is the inverse link function in a canonical generalized linear model for the outcome. Calculate the MLEs, e.g., $(\hat\psi_{j+1,-k}^\ast,\hat\beta_{j+1,-k})$, as well as the observed Fisher information matrix. 

\item Randomly draw a value of the estimated outcome model coefficients $(\widehat\psi_{j+1,-k}^\ast,\widehat\beta_{j+1,-k})$ from their joint sampling distribution, which we denote simply by $G(\cdot)$. In practice, $G(\cdot)$ may be approximated by the (asymptotically-valid) multivariate normal distribution with mean vector $(\hat\psi_{j+1,-k}^\ast,\hat\beta_{j+1,-k})$ and the covariance matrix being the inverse of the observed Fisher information matrix; see e.g., \citet[Section 7.6.2]{wakefield2013bayesian}.
Denote the randomly drawn values by $(\tilde\psi_{j+1,-k}^\ast,\tilde\beta_{j+1,-k})$. Let $\tilde\E(Y|A,L^{j+1, -k})$ denote the {\em perturbed} outcome model obtained by plugging in the randomly drawn values of the outcome model coefficients.

\item Calculate the effect estimator as described in the Appendix, using the observed (exposure) weights and perturbed outcome model. Denote the resulting perturbed estimator by $\tilde \psi_{j+1,k}$.

\een

The difference between $\widehat \psi_{j+1,k}$ and $\tilde \psi_{j+1,k}$ therefore lies in step 2(c), where the former is based on the MLE of the outcome model coefficients whereas the latter is based on a random draw from the estimated coefficients' sampling distribution. 
Calculating the perturbed effect estimators only requires fitting the exposure and outcome models once (to the observed data), which is computationally more efficient than e.g., a nonparametric bootstrap procedure that randomly resamples observations with replacement, and refits both exposure and outcome models.
For computational convenience, we have considered only a perturbed outcome model. In principle, a random draw of the estimated exposure model coefficients from their sampling distribution, denoted by e.g., $\tilde\alpha_{j+1,-k}$, may be made in step 2(a). The weights as defined in \eqref{eq:IPTW_def} can then be calculated by substituting $\tilde\alpha_{j+1,-k}$ for $\alpha_{j+1,-k}$ in the exposure model. The resulting (perturbed) weights are then used to fit the outcome model in step 2(b). 


\section{Fitting natural cubic splines to a trajectory of the exposure effect estimator}\label{sect:extrapolate_naturalcubicsplines}

A natural cubic spline is fitted to each sequence of exposure effect estimators, e.g., $\widehat \psi_{j}, j=1,\ldots,J+1$, with the number of measured covariates adjusted for ($0, \ldots, J$, after excluding the intercept) as a predictor. Natural cubic splines permit flexibly evaluating the trajectory of the exposure effect estimator as unmeasured confounding is reduced one covariate at a time. The predicted value of the effect estimator, had a given number of hypothetical confounders been further adjusted for, can then be linearly extrapolated to.
The extrapolation therefore assumes that the impact of adjusting for hypothetical confounders on the causal effect satisfies the boundary conditions of the fitted spline, e.g., a zero second derivative at the boundary knots. 
For simplicity, natural cubic splines can be fitted with the largest number of (evenly-spaced) interior knots permitted to ensure identifiability of the spline; when there are $J$ confounders, there may be as many as $J-1$ interior knots. This will maximize the flexibility of the fitted spline to more closely capture  changes in the effect estimators as unmeasured confounding is systematically varied. In principle, the number of knots may be selected via cross-validation by fitting multiple natural cubic splines, each with a different number of knots. For example, a natural cubic spline with a given number of knots could be fitted to each perturbed sequence (i.e., the ``training'' data), and the predicted effect estimate in each orbit calculated. The mean squared difference between each prediction and the observed effect (the ``test'' data) could then be calculated for each orbit. The number of knots that minimized the average prediction error across all orbits is then selected. This procedure was used in the illustration with the RHC data, where a natural cubic spline with 53 equally-spaced interior knots was fitted to each (perturbed) sequence of the estimators and endpoints of the 95\% CI. We found that using the maximum number of interior knots in this data resulted in highly fluctuating trajectories that were (much) more variable than the sampling variability represented in the 95\% CI adjusting for the measured covariates. However, such ``out-of-sample'' predictions would likely select a small number of knots, and tend to over-smooth the trajectory and yield poor quality extrapolations. Developing methods for selecting an optimal number of knots is thus deferred to future work.


\section{Simulation studies}\label{sect:sims}
Simulation studies were conducted under different data-generating scenarios to empirically evaluate the ability of the proposal to avoid potential biases due to unmeasured confounding.

\subsection{Study 1}
Data for a single population was generated as follows:
\begin{align*}
L_1,\ldots,L_{p+q} &\underset{i.i.d.}{\sim} {\cal N}(0,1) \\
A &\sim {\rm Bernoulli}\{\expit(\alpha_0 + \sum_{k=1}^{p+q} \alpha_{k} L_k)\} \\
Y &\sim {\rm Bernoulli}\{\expit(\beta_0 + \delta A + \sum_{k=1}^{p+q} \beta_{k} L_k)\}
\end{align*}
We set $\alpha_0=\beta_0=0, \alpha_{k} \sim {\rm Uniform}(-1, 1), \beta_k=\alpha_k, k=1,\ldots,p+q$.
Each confounder had the same strength of influence on the exposure and outcome.
To induce unmeasured confounding, we iteratively removed the $q$ confounders that resulted in the smallest (absolute) changes to the effects between consecutive orbits following the deterministic criterion \eqref{eq:forwardselect_minimin}. These (same) $q$ confounders would thus be regarded as unmeasured (in each observed dataset), so that only $p$ measured confounders remained available for adjustment. We considered values of $p \in \{12,16\}$, and $q \in \{0,4,8\}$.
The true causal effect was zero when $\delta=0$; we considered values of $\delta \in \{0,1\}$.
There were a total of 12 different data-generating scenarios.
For each setting, we first generated a single population of size $N=50000$ (so that sampling variability can be essentially ignored when determining which confounders to be unmeasured), then generated each observed dataset by randomly sampling (without replacement) $n=1000$ individuals from the population. For each observed dataset, we eliminated the measured confounders in turn by minimizing the (squared) differences in the effects between consecutive orbits following \eqref{eq:forwardselect_minimin}. We then constructed the sequence of effect estimators, as well as the 95\% CIs for the effect, indexed by the sequence of nested confounder subsets. For simplicity, we fitted to each sequence a natural cubic spline with the largest number of (evenly-spaced) interior knots permitted to ensure identifiability of the spline; when there are $p$ confounders, there may be as many as $p-1$ interior knots. 
The predicted (inference for the) effect adjusting for $q$ unmeasured confounders was then extrapolated to.
When $q=0$, we extrapolated to two unmeasured confounders to assess the stability of (inference for) the effect.

We simulated 1000 samples for each setting.
We considered point estimates of, and 95\% CIs for, the population average exposure effect $\psi$ that were based on either (i) adjusting for all $p+q$ (un)measured confounders associated with exposure and outcome, or (ii) adjusting for only the $p$ measured confounders, or (iii) the extrapolated predictions.
The empirical mean and standard deviation (across all simulated samples) of all three point estimates are displayed in Table~\ref{table:sim1-res}.
We also assessed the empirical frequency at which each of the 95\% CIs, as well as the uncertainty interval, included the average treatment effect. (Recall that the uncertainty interval for a single observed dataset was the union of all extrapolated 95\% CIs across all perturbed sequences.) 
Adjusting for only the measured confounders yielded only slightly biased point estimates, but resulted in 95\% CIs whose coverage levels were empirically far below the nominal level in the presence of a large number of unmeasured confounders.
The extrapolated predicted effect was similarly slightly biased, but had much greater empirical variability, resulting in the largest empirical mean squared error (MSE) among the three estimates.
When there were many unmeasured confounders, the presence of more measured covariates to learn about the impact of unmeasured confounding from resulted in the uncertainty intervals being more likely to capture the population average effect at (or exceeding) the nominal level of the constituent 95\% CIs in the presence of unmeasured confounding. 

\begin{table}[!ht]
\centering
\begin{tabular}{rrr|rrr|rrr|rrr}
  \hline
& & & \multicolumn{3}{c|}{Point estimates} & \multicolumn{3}{c|}{Empirical MSE (square root)} & \multicolumn{3}{c}{Coverage of 95\% CIs} \\
$q$ & $\psi$ & $p$ & All & Measured & Predicted & All & Measured & Predicted & All & Measured & Predicted \\ 
  \hline
0 & 0.00 & 12 & 0.00 (0.03) & 0.00 (0.03) & 0.00 (0.03) & 0.03 & 0.03 & 0.03 & 0.94 & 0.94 & 0.98 \\ 
  0 & 0.00 & 16 & 0.00 (0.03) & 0.00 (0.03) & 0.00 (0.03) & 0.03 & 0.03 & 0.03 & 0.93 & 0.93 & 0.98 \\ 
  0 & 0.16 & 12 & 0.16 (0.03) & 0.16 (0.03) & 0.15 (0.03) & 0.03 & 0.03 & 0.03 & 0.94 & 0.94 & 0.98 \\ 
  0 & 0.13 & 16 & 0.13 (0.03) & 0.13 (0.03) & 0.13 (0.03) & 0.03 & 0.03 & 0.03 & 0.93 & 0.93 & 0.98 \\ 
  4 & 0.00 & 12 & 0.00 (0.03) & 0.01 (0.03) & 0.00 (0.04) & 0.03 & 0.03 & 0.04 & 0.94 & 0.95 & 0.99 \\ 
  4 & 0.00 & 16 & 0.00 (0.03) & 0.00 (0.03) & -0.01 (0.04) & 0.03 & 0.03 & 0.04 & 0.93 & 0.94 & 0.98 \\ 
  4 & 0.13 & 12 & 0.13 (0.03) & 0.14 (0.03) & 0.13 (0.04) & 0.03 & 0.04 & 0.04 & 0.92 & 0.94 & 0.99 \\ 
  4 & 0.13 & 16 & 0.13 (0.04) & 0.13 (0.04) & 0.12 (0.04) & 0.04 & 0.04 & 0.05 & 0.90 & 0.92 & 0.98 \\ 
  8 & 0.00 & 12 & 0.00 (0.03) & 0.05 (0.03) & -0.05 (0.08) & 0.03 & 0.06 & 0.09 & 0.94 & 0.72 & 0.85 \\ 
  8 & 0.00 & 16 & -0.01 (0.03) & 0.04 (0.03) & -0.04 (0.12) & 0.03 & 0.05 & 0.12 & 0.90 & 0.73 & 0.93 \\ 
  8 & 0.13 & 12 & 0.12 (0.04) & 0.18 (0.04) & 0.08 (0.09) & 0.04 & 0.07 & 0.10 & 0.89 & 0.72 & 0.91 \\ 
  8 & 0.11 & 16 & 0.10 (0.03) & 0.15 (0.04) & 0.08 (0.10) & 0.03 & 0.06 & 0.11 & 0.89 & 0.70 & 0.97 \\ 
   \hline
\end{tabular}
\caption{Results for simulation study 1. 
Point estimates of, and inference for, the average exposure effect $\psi$ were based on either (i) adjusting for all confounders (``All''), or (ii) adjusting for only the measured confounders (``Measured''), or (iii) the extrapolated prediction (``Predicted'').
The empirical MSE was calculated as the sum of the squared bias and the empirical variance of the point estimates.
Coverage for the 95\% CIs was calculated as the empirical proportion of CIs that included the population average exposure effect. For simplicity, the extrapolated uncertainty interval is termed the ``Predicted 95\% CI'' in this table.
Empirical standard errors are in brackets. All results were rounded to two decimal places.\label{table:sim1-res}}
\end{table}

\subsection{Study 2}
In this study, we compare the proposed method with the existing methods of \citet{carnegie2016assessing}, as implemented in the ``treatSens'' package \citep{carnegie2020treatSens}, and of \citet{cinelli2020making}, as implemented in the ``sensemakr'' package \citep{sensemakr2020} in \texttt{R} \citep{R2020}.
We modified the data-generating process of the previous study to fulfill the required assumptions of \citet{carnegie2016assessing}; i.e., exposure was generated as $A \sim \Phi(\alpha_0 + \sum_{k=1}^{p+q} \alpha_{k} L_k)$, where $\Phi(\cdot)$ is the cumulative distribution function of a standard normal variable so that the exposure model employs a ``probit'' link, and outcome $Y \sim {\cal N}(\beta_0 + \delta A + \sum_{k=1}^{p+q} \beta_{k} L_k, p+q)$ was continuous and normally distributed (with variance simply equal to the total variance of all $p+q$ independent confounders with unit variance). \citet{sensemakr2020} require only a continuous outcome so that linear regression models can be fitted, with no exposure model needed. We set $\alpha_0=\beta_0=0, \alpha_{k} \sim {\rm Uniform}(-0.25, 0.25), \beta_k \sim {\rm Uniform}(-4, 4), k=1,\ldots,p+q$.
We considered settings with $p=16, q \in \{0,4,8\}$, and $\delta \in \{0,2\}$.
In addition, to empirically evaluate the biases that can result when the ``probit'' link function in the assumed exposure model is misspecified, we generated data for the exposure under a ``logit'' link function.
There were a total of 12 different data-generating scenarios.

For each setting, we first generated a single population of size $N=50000$ with all $p+q$ confounders, then induced unmeasured confounding by iteratively removing the $q$ confounders that resulted in the smallest (absolute) changes to the effects between consecutive orbits following the deterministic criterion \eqref{eq:forwardselect_minimin}. These (same) $q$ confounders would thus be regarded as unmeasured (in each observed sample), so that only $p$ measured confounders remained available for adjustment. The data for each observed sample was generated by randomly drawing (without replacement) $n=2000$ individuals from the population. For each observed data, we carried out the proposed sensitivity analysis by eliminating the measured confounders in turn to minimize the (squared) differences in the effects between consecutive orbits following \eqref{eq:forwardselect_minimin}. We then constructed the sequence of standardized effect estimators, as well as the 95\% CIs for the effect, indexed by the sequence of nested confounder subsets. 
Natural cubic splines with the maximum number of $p-1$ (equally-spaced interior) knots were fitted to each sequence.
The predicted (inference for the) effect adjusting for $q$ unmeasured confounders was then extrapolated to.
When $q=0$, we extrapolated to two unmeasured confounders to assess the stability of (inference for) the effect.

Both the treatSens and sensemakr methods require specifying values for the sensitivity parameters that encode the separate associations between a single hypothetical unmeasured confounder with the exposure, and with the outcome, respectively, for each individual simulated dataset. For simplicity when implementing these simulation studies, we selected a ``calibrated,'' or ``benchmark,''  value among the coefficient estimates of the $p$ measured confounders in the exposure (or outcome) model fitted to each simulated observed data.
When $q=0$ so that there truly was no unmeasured confounding, we selected the coefficient estimate with the smallest (absolute) magnitude among the $p$ estimated coefficients to assess the impact of incorrectly assuming some (minimal) unmeasured confounding. When $q>0$ so that there truly was unmeasured confounding, we selected the coefficient estimate with the largest (absolute) magnitude among the $p$ estimated coefficients as a proxy for the multiple unmeasured confounders. 
When using the treatSens method, we sampled 500 posterior draws of the exposure effect point estimate, and its standard error estimate, for the given combination of the sensitivity parameter values.
The point estimate was determined as the posterior mean, and the standard error estimate was determined using Rubin's rules as described in \citet{carnegie2016assessing}; 95\% Wald CIs were then constructed using the corresponding standard normal quantiles.
When using the sensemakr method, we selected a bias adjustment that reduces (increases) the absolute value of the estimated coefficient for the exposure in the outcome model if $\psi=0$ ($\psi>0$), even though the true value of $\psi$ is unknown in practice. All other arguments in the sensemakr function were left at their default values.

We simulated 1000 samples for each setting.
We again considered the point estimates of, and 95\% CIs for, the population average exposure effect $\psi$ that were based on either (i) adjusting for all $p+q$ (un)measured confounders associated with exposure and outcome, or (ii) adjusting for only the $p$ measured confounders, or (iii) the extrapolated predictions.
In addition, we considered the predictions using the treatSens and sensemakr methods with their respective calibrated or benchmark values of the sensitivity parameters.
The empirical mean and standard deviation (across all simulated samples) of all five point estimates are displayed in Table~\ref{table:sim2-res}.
Because the treatSens method can fail for a given combination of the calibrated sensitivity parameter values, we included the proportion of simulated datasets where the method returned such an error.
We also assessed the empirical frequency at which each of the 95\% CIs included the average treatment effect.
There were no systematic biases in the point estimates empirically when all the confounders (including those regarded as unmeasured) were adjusted for, regardless of the assumed link function for the exposure model. 
The treatSens method failed in at least 40\% of the simulated datasets; among datasets where the method successfully converged, unbiased point estimates were obtained only when there was truly no exposure effect ($\psi=0$) and no unmeasured confounding ($q=0$). However, the 95\% CIs met the nominal coverage levels only when the probit link was correctly assumed; in all other cases, the CIs failed to capture the true value of $\psi$ in all the simulated samples.
The sensemakr method yielded unbiased estimates when there was truly no exposure effect ($\psi=0$), and positively biased estimates when there was an exposure effect ($\psi>0$). However, the 95\% CIs had empirical coverage levels far below their nominal levels when $\psi>0$ and there was unmeasured confounding ($q>0$).
The extrapolated predicted effects were unbiased (within the empirical standard errors) across all settings. In most settings, the uncertainty intervals had coverage levels that empirically exceeded their nominal levels, except when $q=8$ where the coverage levels fell below their nominal levels empirically.

\begin{table}[!ht]
\centering
\begin{tabular}{rrr|rrrrrr}
  \hline
& & & \multicolumn{5}{c}{Point estimates}  \\  
$q$ & link & $\psi$ & All & Measured & \% fail & treatSens & sensemakr & Predicted \\ 
  \hline
0 & logit & 0.00 & 0.02 (0.18) & 0.02 (0.18) & 0.44 & -0.01 (0.02) & 0.01 (0.17) & 0.02 (0.24) \\ 
  0 & logit & 2.00 & 2.01 (0.19) & 2.01 (0.19) & 0.47 & 0.19 (0.02) & 2.02 (0.19) & 2.01 (0.25) \\ 
  0 & probit & 0.00 & 0.03 (0.20) & 0.03 (0.20) & 0.44 & 0.00 (0.02) & 0.03 (0.18) & 0.03 (0.25) \\ 
  0 & probit & 2.00 & 2.03 (0.20) & 2.03 (0.20) & 0.43 & 0.20 (0.02) & 2.05 (0.19) & 2.03 (0.24) \\ 
  4 & logit & 0.00 & -0.02 (0.20) & 0.11 (0.29) & 0.60 & -3.37 (0.97) & -0.06 (0.37) & 0.16 (0.53) \\ 
  4 & logit & 2.00 & 1.99 (0.20) & 2.12 (0.28) & 0.61 & -3.28 (0.83) & 2.60 (0.36) & 2.17 (0.52) \\ 
  4 & probit & 0.00 & -0.03 (0.23) & 0.16 (0.33) & 0.60 & -3.44 (0.84) & -0.19 (0.60) & 0.32 (0.66) \\ 
  4 & probit & 2.00 & 1.98 (0.24) & 2.16 (0.32) & 0.59 & -3.23 (0.91) & 2.94 (0.44) & 2.32 (0.67) \\ 
  8 & logit & 0.00 & 0.06 (0.23) & -0.02 (0.32) & 0.64 & -1.66 (1.38) & -0.01 (0.33) & 0.21 (1.15) \\ 
  8 & logit & 2.00 & 2.04 (0.22) & 1.97 (0.33) & 0.61 & -1.63 (1.42) & 2.35 (0.41) & 2.25 (1.11) \\ 
  8 & probit & 0.00 & 0.04 (0.29) & -0.08 (0.38) & 0.66 & -1.92 (1.46) & 0.03 (0.52) & 0.59 (1.56) \\ 
  8 & probit & 2.00 & 2.03 (0.27) & 1.88 (0.37) & 0.64 & -1.70 (1.45) & 2.47 (0.49) & 2.62 (1.48) \\ 
   \hline
\end{tabular}
\begin{tabular}{rrr|rrrrr}
  \hline
& & & \multicolumn{5}{c}{Empirical MSE (square root)}  \\
$q$ & link & $\psi$ & All & Measured & treatSens & sensemakr & Predicted \\
  \hline
0 & logit & 0.00 & 0.18 & 0.18 & 0.02 & 0.17 & 0.24 \\ 
  0 & logit & 2.00 & 0.19 & 0.19 & 1.81 & 0.19 & 0.25 \\ 
  0 & probit & 0.00 & 0.20 & 0.20 & 0.02 & 0.18 & 0.26 \\ 
  0 & probit & 2.00 & 0.21 & 0.21 & 1.80 & 0.20 & 0.24 \\ 
  4 & logit & 0.00 & 0.21 & 0.31 & 3.50 & 0.38 & 0.56 \\ 
  4 & logit & 2.00 & 0.21 & 0.31 & 5.35 & 0.70 & 0.55 \\ 
  4 & probit & 0.00 & 0.23 & 0.37 & 3.54 & 0.63 & 0.73 \\ 
  4 & probit & 2.00 & 0.24 & 0.36 & 5.31 & 1.04 & 0.75 \\ 
  8 & logit & 0.00 & 0.24 & 0.32 & 2.16 & 0.33 & 1.17 \\ 
  8 & logit & 2.00 & 0.23 & 0.33 & 3.89 & 0.54 & 1.14 \\ 
  8 & probit & 0.00 & 0.29 & 0.39 & 2.41 & 0.52 & 1.67 \\ 
  8 & probit & 2.00 & 0.27 & 0.39 & 3.97 & 0.67 & 1.61 \\ 
   \hline 
& & & \multicolumn{5}{c}{Coverage of 95\% CIs} \\
$q$ & link & $\psi$ & All & Measured & treatSens & sensemakr & Predicted \\
  \hline
0 & logit & 0.00 & 0.96 & 0.96 & 0.00 & 0.96 & 0.99 \\ 
  0 & logit & 2.00 & 0.95 & 0.95 & 0.00 & 0.94 & 0.98 \\ 
  0 & probit & 0.00 & 0.95 & 0.95 & 1.00 & 0.95 & 0.98 \\ 
  0 & probit & 2.00 & 0.95 & 0.95 & 0.00 & 0.95 & 0.98 \\ 
  4 & logit & 0.00 & 0.95 & 0.93 & 0.00 & 0.76 & 0.98 \\ 
  4 & logit & 2.00 & 0.96 & 0.94 & 0.00 & 0.35 & 0.97 \\ 
  4 & probit & 0.00 & 0.94 & 0.91 & 0.00 & 0.43 & 0.95 \\ 
  4 & probit & 2.00 & 0.94 & 0.92 & 0.00 & 0.16 & 0.94 \\ 
  8 & logit & 0.00 & 0.94 & 0.96 & 0.00 & 0.91 & 0.81 \\ 
  8 & logit & 2.00 & 0.96 & 0.96 & 0.00 & 0.69 & 0.82 \\ 
  8 & probit & 0.00 & 0.93 & 0.94 & 0.00 & 0.79 & 0.84 \\ 
  8 & probit & 2.00 & 0.95 & 0.95 & 0.00 & 0.63 & 0.86 \\ 
   \hline
\end{tabular}
\caption{Results for simulation study 2. 
Point estimates of, and inference for, the average exposure effect $\psi$ were based on either (i) adjusting for all confounders (``All''), or (ii) adjusting for only the measured confounders (``Measured''), or (iii) the extrapolated prediction (``Predicted''), or (iv) the treatSens method assuming calibrated values of the sensitivity parameters, or (v) the sensemakr method assuming benchmark values of the sensitivity parameters.
The proportion of simulated datasets where the treatSens method failed is labelled as ``\% fail.''
The empirical MSE was calculated as the sum of the squared bias and the empirical variance of the point estimates.
Coverage for the 95\% CIs was calculated as the empirical proportion of CIs that included the population average exposure effect.
Empirical standard errors are in brackets. All results were rounded to two decimal places.\label{table:sim2-res}}
\end{table}

\clearpage

\section{AIDS Clinical Trials Group Study 175}\label{sect:ACTG175}

The `ACTG175' dataset was from an AIDS randomized clinical trial, and was distributed as part of the \texttt{speff2trial} package via the Comprehensive R Archive Network (\url{https://CRAN.R-project.org/package=speff2trial}). The trial compared monotherapy using either zidovudine or didanosine alone with combination therapy using either zidovudine and didanosine, or zidovudine and zalcitabine, in adults infected with the human immunodeficiency virus type I whose CD4 T cell counts were between 200 and 500 per cubic millimeter. Exposure was (re)coded as $A=0$ for therapy using either zidovudine or didanosine only, and $A=1$ for therapies combining zidovudine and either didanosine or zalcitabine. A binary outcome was defined based on whether a participant's CD4 T cell count at $96\pm 5$ weeks was greater than 250 or not. The full dataset contained 2139 participants with 17 measured (putative) confounders, but we only considered a reduced dataset with 1342 participants having complete data so that an exposure model with all covariates could be fitted. In addition, one covariate (prior zidovudine use) that was singular in the reduced dataset was dropped.

\begin{figure}[!ht]
    \centering
\includegraphics[width=\linewidth,keepaspectratio]{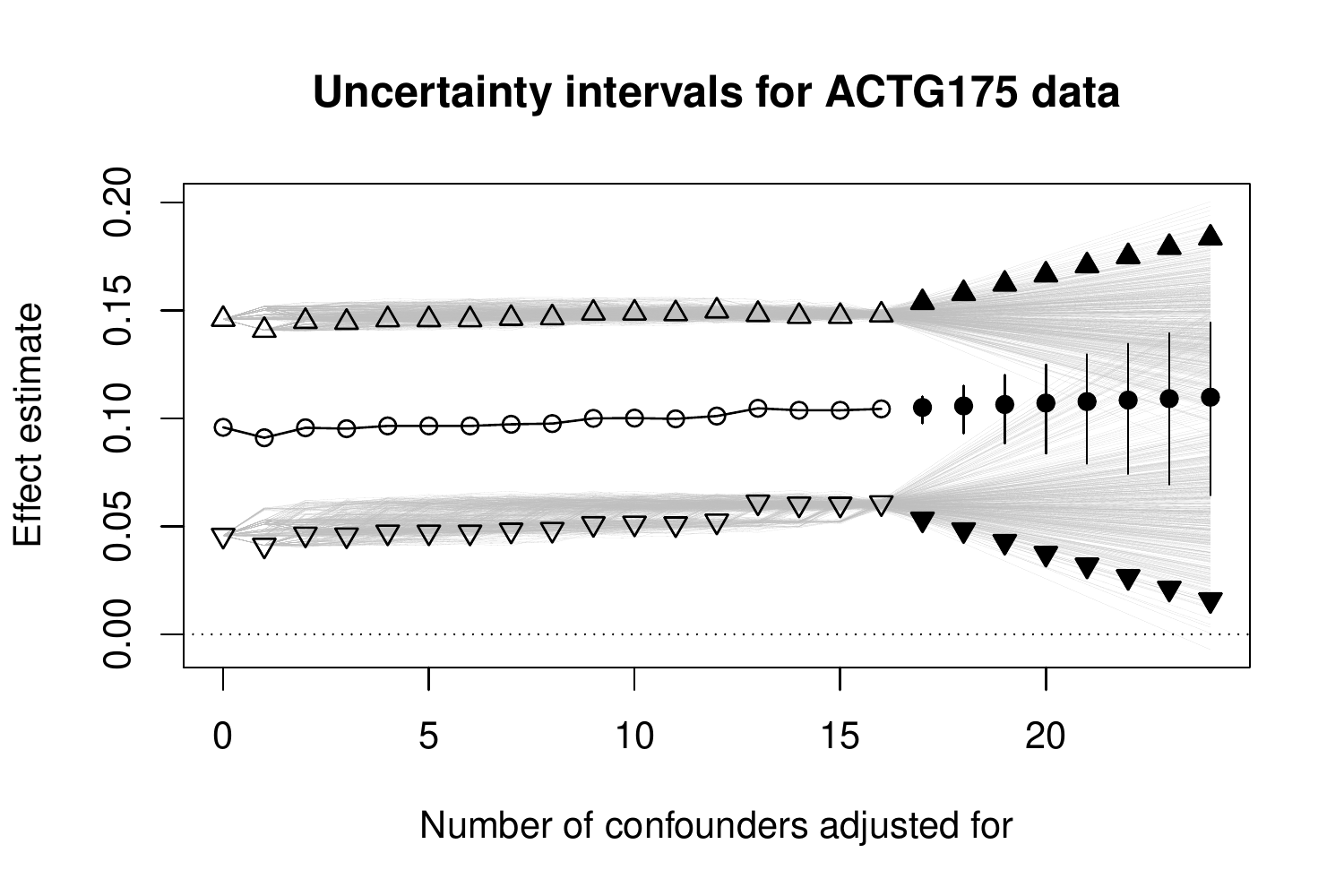}
    \caption{Estimators of the average exposure effect from each orbit, for the ACTG175 data. 
Circles indicate the point estimates, and (inverted) triangles indicate the (lower) upper endpoint of the 95\% CI.
Empty points indicate the sequence of effect estimates constructed by eliminating a single measured confounder from adjustment in turn. The curve corresponds to a fitted natural cubic spline.
Filled points indicate the (extrapolated) predictions after adjusting for $q$ additional unmeasured confounders. 
Each thin grey line corresponds to the natural cubic spline fitted to a perturbed sequence. 
The vertical lines indicate the symmetric 95 percentiles of the predicted effects across all $B$ perturbations.
    \label{fig:plot-speff2}}
    \end{figure}

The sequence of effect estimators constructed by eliminating covariates one at a time following the criterion in \eqref{eq:forwardselect_minimin} is plotted (as empty circles) in Figure~\ref{fig:plot-speff2}.  
Because this was a randomized controlled trial, confounding of the exposure-outcome relation was unlikely, but could have been induced by the exclusion of incomplete observations. Adjusting for different covariates did not greatly affect the exposure effect estimates, as shown by the relatively stable trajectory across different adjustment sets: across all the orbits, the estimated effect was about 0.10, with a 95\% CI approximately between 0.05 and 0.15.
We calculated $B=500$ perturbed sequences of the estimators and their corresponding 95\% CIs.
Across all the perturbations, the 95\% CIs for the exposure effect excluded zero, even when (all) the measured covariates had been intentionally eliminated from adjustment.
A natural cubic spline with the maximum number of 15 equally-spaced interior knots was fitted to each (perturbed) sequence of the estimators and endpoints of the 95\% CI.
The predicted effects, and uncertainty intervals, adjusting for between one and eight (half the number of measured covariates) hypothetical confounders were then extrapolated to. We trimmed 5\% of the extreme endpoints of the perturbed CIs to construct the uncertainty intervals.
Further adjustment for hypothetical unmeasured confounding appeared unlikely to yield an effect estimate that would be significantly indistinguishable from zero: the uncertainty interval excluded zero even after adjusting for seven additional hypothetical (unmeasured) confounders, as indicated by the filled triangles being above zero in Figure~\ref{fig:plot-speff2}. 
The results suggested that statistical inference for the effect was unlikely to change (from adjusting for only the measured confounders).
The conclusion that there was a (statistically significant) positive average exposure effect of combination therapy on CD4 T cell count thus appeared to be insensitive to (hypothetical) unmeasured confounding.

\section{Existing methods for sensitivity analysis}\label{sect:review_existing}
Existing methods for assessing sensitivity to unmeasured confounding typically judge the (minimum) extent that adjusting for a hypothetical scalar confounder, conditional on the measured confounders, affects inference about the causal effect. 
A widely-adopted approach focuses on characterizing the (strength and direction of) association between the hypothetical confounder and the exposure, either as a regression coefficient or the odds ratio of receiving exposure, as a sensitivity parameter \citep{Rosenbaum1987,Rosenbaum:2002,tan2006distributional,zhao2019sensitivity,rosenbaum2020conditional}.
Another approach further incorporates the (strength and direction of) association between the hypothetical confounder and the outcome as an additional sensitivity parameter \citep{rosenbaum1983assessing,Lin1998,imbens2003}. 
Recent methods adopting the latter approach include \citet{VanderWeele2011}, \citet{richardson2014nonparametric}, \citet{carnegie2016assessing}, \citet{ding2016sensitivity}, \citet{dorie2016flexible}, \citet{vanderweele2017sensitivity}, \citet{genback2019causal}, \citet{zhang2019semiparametric}, \citet{cinelli2020making}, \citet{veitch2020sense}, among many  others.
We refer readers to \citet{liu2013introduction} and \citet{carnegie2016assessing} for an overview of other approaches, including methods that require restrictive assumptions about the parametric form of the outcome model that encodes the exposure effect, and to \citet{veitch2020sense} for an overview of other methods that do not explicitly consider a hypothetical unmeasured confounder and involve other (more complex) sensitivity parameters.
Methods using sensitivity parameters that encode the hypothetical confounder's separate associations with exposure and outcome typically require first specifying a discrete grid of values for the parameters, then estimating the causal effect after adjusting for the hypothetical confounder under different fixed parameter values.
Graphical tools, such as bivariate sensitivity contour plots \citep{carnegie2016assessing,ding2016sensitivity,dorie2016flexible,vanderweele2017sensitivity,veitch2020sense} that visualize different combinations of how strongly the hypothetical confounder must be (simultaneously) associated with exposure and outcome for the conclusions to be affected, can be especially useful for applied researchers.

But sensitivity analyses predicated on such sensitivity parameters suffer from a number of shortcomings. First, users of grid-based approaches must explicitly state the granularity, range, and even signs, for plausible values of the separate (conditional) associations between the hypothetical confounder and the exposure, and the outcome. Reference values are sometimes estimated or postulated based on the measured confounders' (conditional) associations with exposure (and outcome) via an empirical ``calibration.'' A specific measured confounder is selected as a reference based on e.g., its (standardized) estimated regression coefficient \citep{carnegie2016assessing}; or its marginal effect size on the outcome \citep{dorie2016flexible}; or its relative risk on the (binary) outcome conditional on other measured confounders \citep{vanderweele2017sensitivity}; or its contribution to the variation in the outcome \citep{veitch2020sense}.
Bayesian sensitivity analyses for unmeasured confounding \citep{greenland2003impact,McCandless2007,mccandless2017comparison,groenwold2018adjustment} avoid such user calibration through the application of Bayes Theorem to obtain a posterior distribution for the sensitivity parameters, but require more restrictive parametric assumptions about the outcome model.
Second, with the exception of \citet{bonvini2019sensitivity}, \citet{zhang2019semiparametric} and \citet{cinelli2020making}, most sensitivity analyses that evaluate the associations between a hypothetical confounder with exposure, or outcome, or both, (implicitly) require certain assumptions about whether the confounder is continuous or dichotomous, or the distribution of the confounder, or whether the bias is away or toward the null.
Third and most crucially, sensitivity analyses based on such sensitivity parameters can potentially lead to scientifically meaningless and logically incoherent conclusions \citep{robins2002covariance}. 
Consider the following simple example where the (true) probability of receiving a binary exposure ($A=1$) depends on a logistic model with main effects for three confounders, $L_1, L_2$, and $U$; e.g., 
$\Pr(A=1|L_1,L_2,U)=\expit(\beta_0 + \beta_1 L_1 + \beta_2 L_2 + \Gamma U)$, where $\expit(x) = \exp(x)/\{1+\exp(x)\}$. 
The interpretation of the conditional odds ratio of being exposed due to $L_1$ under the true model, $\exp(\beta_1)$, depends on the values of the other confounders $L_2$ and $U$, even in the absence of any interactions. 
Suppose that the researcher could only collect data on $L_1$ and $L_2$, so that $U$ is left unmeasured, and is thus restricted to fitting a different treatment model, e.g., $\Pr(A=1|L_1,L_2)=\expit(\beta_0^\ast + \beta_1^\ast L_1 + \beta_2^\ast L_2)$. The interpretation of $\exp(\beta_1^\ast)$ under the fitted model thus depends on only  the value of $L_2$. It follows that the interpretation of the sensitivity parameter, $\exp(\Gamma)$, is incompatible with both $\exp(\beta_1)$ and $\exp(\beta_1^\ast)$ due to {\em non-collapsibility} \citep{greenland1999confounding}. Of course, one may consider a different treatment model instead, e.g., $\Pr(A=1|L_2,U)=\expit(\beta_0^{\ast\ast} + \beta_2^{\ast\ast} L_2 + \Gamma^\ast U)$, so that the interpretation of the sensitivity parameter $\exp(\Gamma^\ast)$ that encodes the conditional odds ratio due to $U$ given $L_2$ is seemingly comparable with $\exp(\beta_1^\ast)$. But there is no guarantee that the range of plausible values for $\Gamma$ will necessarily be narrower than that for $\Gamma^\ast$, even though the former is based on an exposure model that adjusts for both $L_1$ and $L_2$ thereby reducing the extent of biases due to unmeasured confounding. It can therefore be difficult (or impossible) to conceptualize sensitivity parameters that retain the same meaningful interpretation regardless of the measured confounders being adjusted for.

\end{document}